\documentclass[11pt,a4paper]{article}

\usepackage[margin=2 cm,twoside]{geometry}
\usepackage[english]{babel}
\usepackage[utf8]{inputenc}
\usepackage[T1]{fontenc}
\usepackage{siunitx} %SI units
\usepackage{fancyhdr} %Laver fine header
\usepackage{lastpage} % Referere til sidste side
\usepackage[notlof,notlot,nottoc]{tocbibind}

\usepackage[font=footnotesize,labelfont=bf]{caption}

\usepackage{float}
\usepackage{amsthm} %Thereom package, for creating, numbering, styling and making proofs.
\usepackage{amsmath} 
\usepackage{amssymb} %Makes symbols such as: \barwedge, \Box and \Diamond.
\usepackage{graphicx} 
\usepackage{color} %Change color on text, pages and possibly more
\usepackage{subcaption}
\usepackage{physics} %Able to easily make bra and ket notation e.g. \ket{\psi}
\usepackage{tikzscale}
\usepackage{cite}
\usepackage{url} %Indsætte links
\usepackage{wrapfig}
\usepackage[toc,page]{appendix}
\usepackage[hidelinks]{hyperref}

\newlength\figH
\newlength\figW
\usepackage {tikz}
\usetikzlibrary{positioning}
\usepackage{pgfplots} %Create your own graphs directly in LaTeX.
\pgfplotsset{compat=1.12}

\newcommand{\ignore}[1]{}

\begin{document}

\noindent{\huge\bfseries Highly efficient, tunable, electro-optic metasurfaces based on quasi-bound states in the continuum} \\

\noindent{\large Christopher Damgaard-Carstensen\textsuperscript{1}, Torgom Yezekyan\textsuperscript{2}, Mark L. Brongersma\textsuperscript{3}, \\and Sergey I. Bozhevolnyi\textsuperscript{1} \\
	
\noindent\textsuperscript{1}\textit{Centre for Nano Optics, University of Southern Denmark, Campusvej 55, DK-5230 Odense M, Denmark} \\
\textsuperscript{2}\textit{POLIMA — Center for Polariton-driven Light–Matter Interactions, University of Southern Denmark, Campusvej 55, DK-5230 Odense M, Denmark} \\
\textsuperscript{3}\textit{Geballe Laboratory for Advanced Materials, Stanford University, Stanford, California 94305, United States} }

\section*{Abstract}
Ultrafast and highly efficient dynamic optical metasurfaces enabling truly spatiotemporal control over optical radiation are poised to revolutionize modern optics and photonics, but their practical realization remains elusive. In this work, we demonstrate highly efficient electro-optic metasurfaces based on quasi-bound states in the continuum (qBIC) operating in reflection that are amenable for ultrafast operation and thereby spatiotemporal control over reflected optical fields. The material configuration consists of a lithium niobate thin film sandwiched between an optically thick gold back-reflector and a grating of gold nanoridges also functioning as control electrodes. Metasurfaces for optical free-space intensity modulation are designed by utilizing the electro-optic Pockels effect in combination with an ultra-narrow qBIC resonance, whose wavelength can be finely tuned by varying the angle of light incidence. The fabricated electro-optic metasurfaces operate at telecom wavelengths with the modulation depth reaching 95 \% (modulating thereby 35 \% of the total incident power) for a bias voltage of $\pm$30 V within the electrical bandwidth of 125 MHz. Leveraging the highly angle-dependent qBIC resonance realized, we demonstrate electrically tunable phase contrast imaging using the fabricated metasurface. Moreover, given the potential bandwidth of 39 GHz estimated for the metasurface pixel size of 22 µm, the demonstrated electro-optic metasurfaces promise successful realization of unique optical functions, such as harmonic beam steering and spatiotemporal shaping as well as nonreciprocal operation.

\section*{Introduction}
In recent years, research into dynamic optical metasurfaces, consisting of controllable planar arrays of subwavelength elements, has increased significantly \cite{Thomaschewski2022,Sinatkas2021,Shalaginov2020,Che2020}. The further improvement of dynamic metasurfaces will unlock many new application areas in technologies such as spatial light modulators (SLMs) \cite{Smolyaninov2019,BeneaChelmus2021,Park2020_SLM}, light detection and ranging (LIDAR) \cite{Park2020_SLM,Schwarz2010}, computational imaging and sensing \cite{Jung2018}, and virtual and augmented reality systems \cite{Gopakumar2024,Joo2022}. The main drawback of using metasurfaces for efficient control of radiation is that the interaction length is severely limited by the fundamentally thin nature of metasurfaces. This can be remedied by utilizing materials and configurations that allow for large refractive index variations, such as structural reconfigurations \cite{She2018,Li2016}, phase-change materials \cite{Wang2021,Zhang2021,Abdollahramezani2022}, materials with large thermo-optic effects \cite{Sharma2020,Rahmani2017}, or MEMS configurations \cite{Meng2021,Meng2022,Arbabi2018}. The disadvantage of all these materials and configurations is their inherently slow switching speeds. Recently reported electrically tunable ITO-based metasurfaces, although demonstrating promising developments, were still limited to a few MHz in bandwidth in their experimental realizations \cite{Shirmanesh2020,Park2015,Sisler2024}. 

The linear electro-optic (Pockels) effect, which is found in several ferroelectric media without centrosymmetry like electro-optic polymers \cite{BeneaChelmus2022,Zhang2023}, lead zirconate titanate (PZT) \cite{Alexander2018,Yezekyan2024}, and lithium niobate (LN) \cite{Thomaschewski2020,Liu2024,DamgaardCarstensen2022}, offers inherently fast electrical control of refractive index changes. LN provides large electro-optic coefficients ($r_{33}$ = 31.45 pm/V for the extraordinary polarization and $r_{13}$ = 10.12 for the ordinary \cite{Jazbinek2002}), a high Curie temperature ($\sim\,$1200 \textsuperscript{o}C), a wide transparency range (0.35-4.5 µm), and great mechanical and chemical stability \cite{Weis1985}, all of which makes it an attractive platform for dynamic optical components \cite{Zhang2021_LNreview}. 
The main challenge faced when implementing electro-optic metasurfaces is that very small refractive index changes due to the Pockels effect in combination with the aforementioned severe limitations in the interaction length result in weak modulation of optical fields \cite{Gao2021,Weigand2021}. To circumvent this formidable challenge, one should exploit resonant configurations that would increase the effective interaction length either by multiple reflections across the metasurface layer in Fabry-Perot resonances \cite{DamgaardCarstensen2021} or by nonlocal interaction with the waveguide modes propagating along the metasurface layer \cite{DamgaardCarstensen2023}.

Here, we make the next important step in the latter direction by adding band gap effects to the nonlocal interaction with the waveguide modes and making thereby use of quasi-bound states in the continuum (qBIC) that result in ultra-narrow qBIC resonances \cite{Yezekyan2024_qBIC}. We investigate the occurrence of qBIC resonances in the LN electro-optic metasurface platform \cite{DamgaardCarstensen2021} with symmetric and asymmetric gratings of gold electrodes, being inspired by extremely high Q resonances (> 2$\times$10\textsuperscript{5}) realized with all-dielectric configurations \cite{Cotrufo2024,Huang2023}.
Through the use of numerical simulations and extensive experiments, we utilize the presence of a qBIC resonance to design, fabricate, and experimentally characterize a metasurface for free-space intensity modulation. The fabricated electro-optic metasurfaces operate at telecom wavelengths with the modulation depth reaching 95 \% combined with an absolute modulation of 35 \% for a bias voltage of $\pm$30 V within the electrical bandwidth of 125 MHz. Furthermore, the operation wavelength of our metasurface can be tuned by changing the angle of incidence, and we show that our device maintains a modulation depth of 30-40 \%  within a wavelength range of more than 30 nm by changing the angle of incidence by 1\textsuperscript{o}. Leveraging the highly angle-dependent qBIC resonance realized, we demonstrate electrically tunable phase contrast imaging using the fabricated metasurface. Moreover, given the potential bandwidth of 39 GHz estimated for the metasurface pixel size of 22 µm, the demonstrated electro-optic metasurfaces promise successful realization of unique optical functions, such as harmonic beam steering and spatiotemporal shaping as well as nonreciprocal operation

\section*{Results}

\begin{figure}[tb]
	\centering
	\includegraphics{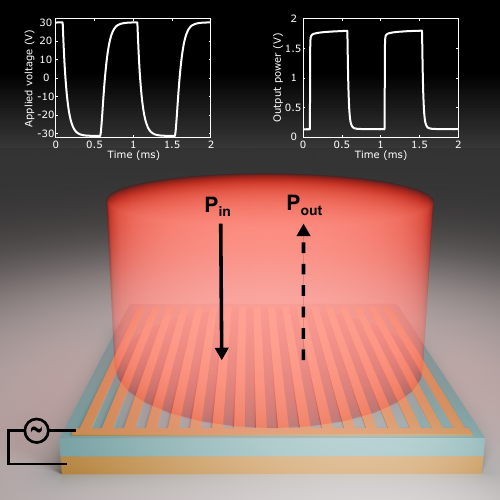}
	\caption{3D rendering of our tunable qBIC metasurface with an incident continuous beam and a modulated outgoing beam. Top plots show the experimentally applied electrical square-shaped signal (left plot) and the corresponding measured modulated optical output (right plot). }
	\label{figIntro}
\end{figure}

\subsection*{Analysis of BIC and qBIC occurrence}
We begin our investigation with numerical analysis of the formation of qBIC and BIC (bound states in the continuum) resonances in a planar waveguide, consisting of a thin $z$-cut LN layer (880 nm) on an optically thick gold platform (300 nm) bonded together with a thin chromium layer (10 nm) and topped with a grating of gold ridges needed to introduce electro-optic control of waveguide modes. This waveguide configuration (without grating electrodes) is commercially available as thin-film LN wafers from NANOLN and was used in our experiments. The grating-assisted excitation of waveguide modes at telecom wavelengths for normal light incidence by using symmetric and asymmetric dielectric grating structures as well as the associated occurrence of BIC and qBIC resonances is well elucidated in our previous work \cite{Yezekyan2024_qBIC}. Use of gold grating ridges instead of dielectric ridges along with the presence of a chromium-gold back reflector introduces (absorption) losses into an otherwise lossless configuration considered previously, influencing significantly the dynamics of resonance processes. The essential physics remains however the same: the resonant excitation of waveguide modes under normal light incidence, taking place at the first diffraction order, is intrinsically linked with the occurrence of Bragg reflection of counter-propagating waveguide modes and results thereby in the formation of a distributed Bragg resonator (DBR) along with BIC and qBIC resonances \cite{Yezekyan2024_qBIC}. 

\begin{figure}[tb]
	\centering
	\includegraphics{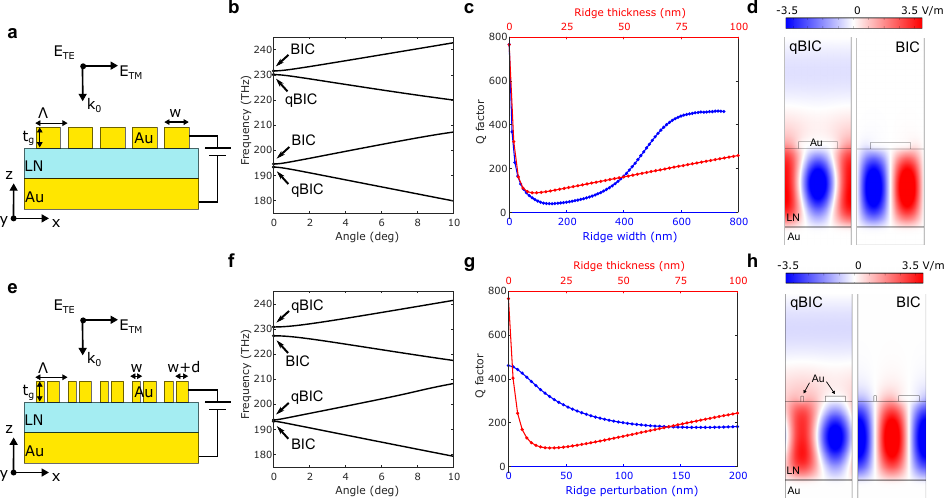}
	\caption{Investigation of modes in the geometry and calculation of Q factor. \textbf{a-d} are for the symmetric grating, and \textbf{e-h} are for the asymmetric grating. \textbf{a,e} Schematic drawing of a metasurface consisting of a \textbf{a} symmetric and \textbf{e} asymmetric gold grating on top of 880 nm of LN on a gold substrate. \textbf{b,f} Simulated dispersion curves. \textbf{c,g} Simulated Q factor vs. \textbf{c} ridge width (lower axis, blue curve) and ridge thickness (upper axis, red curve) and \textbf{g} ridge perturbation (lower axis, blue curve) and ridge thickness (upper axis, red curve) for the fundamental qBIC mode branch under normal incidence. \textbf{d,h} Simulated TE field profiles for the fundamental qBIC (left) and BIC (right) modes for an incident field of 1 V/m. \textbf{a-d} The structure parameters are $\Lambda$ = 750 nm, $w$ = 445 nm and $t_g$ = 75 nm. \textbf{e-h} The structure parameters are $\Lambda$ = 750 nm, $w$ = 25 nm, $d$ = 200 nm, and $t_g$ = 60 nm}
	\label{fig1}
\end{figure}

Let us first consider symmetric thin gold grating couplers on LN (Figure \ref{figIntro}). The efficient excitation of two counter-propagating waveguide modes (excitation in the first diffraction order), means that the waveguide mode wavelength ($\lambda_{eff}$) should be equal to the grating period ($\Lambda$): $\lambda_{eff} = \Lambda$ (Figure \ref{fig1}a). The condition for DBR realization, on the other hand, is $2\Lambda = m\lambda_{eff}$, where $m$ is the order of Bragg reflection. The only way to satisfy both conditions, namely the coupling in the first diffraction order and DBR, is to set $m$ = 2 (i.e., second order Bragg reflection). Here, we are interested in a low-loss fundamental TE mode, which in the desired telecom wavelength range ($\sim\;$1550 nm) has an effective mode refractive index of 2.08 leading to a grating period of $\Lambda$ = 750 nm. The dispersion curves of waveguide modes are depicted in Figure \ref{fig1}b, the qBIC is appearing in the lower energy side of the band gap, with the mode energy mainly concentrated beneath the gold stripe  (Figure \ref{fig1}d). The dependence of the qBIC Q factor on the ridge width (for a fixed height) and height (for a fixed width) is presented in Figure \ref{fig1}c. Here one can note relatively large Q factors for very small ridge widths which is due to less absorption and weak coupling, further from some point, the increase in the ridge width is leading to an increase of the Q factor since the strength of the scattering (coupling) is again reduced. Even though the BIC itself is not accessible with a plane wave illumination, a slight shift from the normal incidence is leading to another important regime, so called “near-BIC”, with characteristic high Q factors compared to qBIC \cite{Azzam2018}. Hence an introduced small angle of incidence is leading to a slight shift of the qBIC resonance position, and occurrence of a new resonance associated with the BIC (Figure \ref{fig1}b), where the distance between these two resonances is linked to the band gap. These features are also evident in reflection spectra of the configuration (Figure \ref{fig2}a and Supplementary Figure S1). Initially at the normal incidence, the qBIC resonance (reflection minima) is at 1551 nm, while with an incidence angle of 1\textsuperscript{o}, it is shifted to 1558 nm, and the near-BIC is excited at a shorter wavelength of 1535 nm (larger energy). Further increase of the angle of incidence will lead to consequent shift of the resonances, making it tunable within certain limits.

Similar regimes can be obtained also with asymmetric grating couplers, i.e., when widths of every M\textsuperscript{th} ridge is altered. Here we consider a structure with doubled period, i.e., every second ridge is perturbed, $\Lambda = 2\lambda_0$, where $\Lambda$ is the supercell period, and $\lambda_0$ is the period of individual ridges (Figure \ref{fig1}e). Here we follow the same logic for the coupling as for the symmetric case, hence $\lambda_{eff} = \Lambda$ holds, however, the DBR in this case is formed from ridges with period of $\lambda_0$, and consequently the DBR condition will be $2\lambda_0 = m\lambda_{eff}$. Again, combining the two conditions (coupling to the first diffraction order and DBR) one will get $m$ = 1, meaning that in the case of the asymmetric grating one deals with the first-order Bragg reflection, when considering the unperturbed grating structure. The corresponding dispersion curves for this configuration are shown in Figure \ref{fig1}f. Unlike the case with symmetric structure, here the qBICs are formed in the higher energy side of the band gap, which however is still consistent with previous case, since the mode energy is again concentrated beneath the gold stripes (Figure \ref{fig1}h). The dependence of the qBIC Q factor on the ridge perturbation (i.e., variation of every second ridge width for a fixed height) and height (for a fixed width) is presented in Figure \ref{fig1}g. Here one should note that the absence of any asymmetry results in resonances with large Q factors, since the mode excitation in the first order vanishes. 
Overall, both configurations have similar features in terms of Q factors and one can implement these structures to make use of narrow resonances.

\subsection*{Design and optimization of metasurface}
By connecting the gold ridges in one end, thus forming a previously used comb-like grating configuration \cite{DamgaardCarstensen2022,DamgaardCarstensen2023}, the gold grating and back-reflector can function as integrated metal electrodes for efficient application of bias signals. The linear electro-optic Pockels effect gives rise to a change in the refractive index of LN, when a bias voltage is applied over the thin film. The change in refractive index is given by the first-order derivation \cite{Thomaschewski2022}: 
\begin{equation}
	|\Delta n| \simeq \frac{1}{2}n^3r_{hk} \frac{V}{d}
\end{equation}
where $n$ is the refractive index without bias voltage, $r_{hk}$ is the relevant Pockels coefficient, $d$ is the thickness of the LN thin-film, and V is the applied bias voltage. 

The design procedure follows a similar reasoning to our previous work \cite{DamgaardCarstensen2023}. To achieve strong and efficient modulation one should look for a sharp and deep resonance. The large Q factors of qBICs offer a sharp resonance, and by simultaneously assuring that scattering and absorption losses are equal, the resonance can be operated at critical coupling, which results in complete radiation absorption \cite{Wu2011}. We optimize our design through numerical simulations of one unit cell using a model with periodic boundary conditions, i.e., assuming an infinite grating. As previously mentioned, the grating period was fixed based on the desired wavelength range. Afterwards, we sweep the ridge width, $w$, thickness, $t_g$, and perturbation, $d$, for the asymmetric grating to achieve critical coupling. The optimized design parameters are: $\Lambda$ = 750 nm, $t_g$ = 75 nm, $w$ = 445 nm for the symmetric grating and $\Lambda$ = 750 nm, $t_g$ = 60 nm, $w$ = 25 nm, $d$ = 200 nm for the asymmetric grating.

\begin{figure}[tb]
	\centering
	\includegraphics{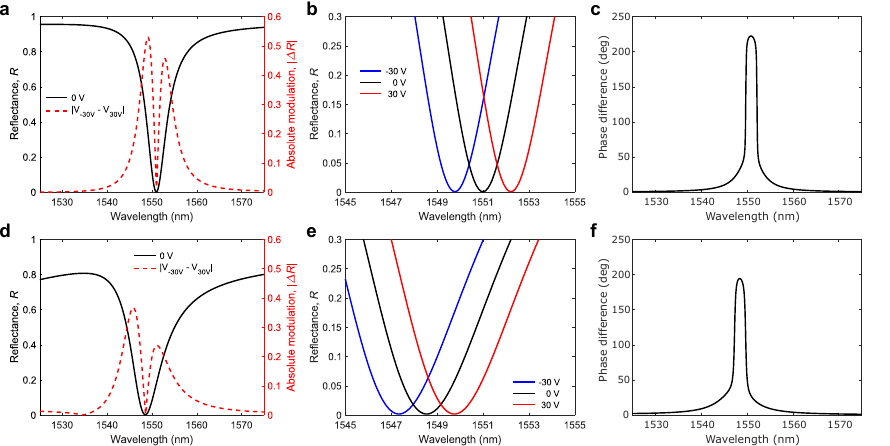}
	\caption{Simulated performance of tunable qBIC metasurface. \textbf{a-c} represent simulations for the symmetric grating, and \textbf{d-f} represent simulations for the asymmetric grating. \textbf{a,d} Reflectance (left axis, black line) and absolute modulation (right axis, red dashed line) vs. wavelength for bias voltages of 0 V and $\pm$30 V, respectively. \textbf{b,e} Zoomed in view of reflectance vs. wavelength for bias voltages of -30 V (blue line), 0 V (black line), and 30 V (red line). \textbf{c,f} Achievable phase difference vs. wavelength when switching from -30 V to 30 V. }
	\label{fig2}
\end{figure}

Two of the primary metrics for estimating performance are the absolute modulation (i.e., the amount of total incident power that is modulated), and the modulation depth or relative modulation (i.e., the amount of reflected power that is modulated). We apply a bias voltage of $\pm$30 V to calculate the absolute modulation $|\Delta R|$, where $R$ is reflectance, and the modulation depth, which is defined as $1-(R_{min}(\lambda)/R_{max}(\lambda))$, where $R_{min}(\lambda)$ and $R_{max}(\lambda)$ are minimum and maximum reflectance values, respectively. 
The simulated spectrum and modulation plots show that critical coupling is achieved (complete radiation absorption), and as a consequence thereof the expected modulation depth of 100 \% is achieved (Figure \ref{fig2}a,d and Supplementary Figure S2). Furthermore, our design shows an absolute modulation of 50 \% and a phase modulation of up to 220\textsuperscript{o} for the symmetric grating (Figure \ref{fig2}a,c) and an absolute modulation of 35 \% and a phase modulation of 200\textsuperscript{o} for the asymmetric grating (Figure \ref{fig2}d,f). The ability to obtain ultrafast phase modulation larger than $\pi$ opens a new avenue towards true spatiotemporal applications \cite{Shaltout2019}. 
A zoom-in view of the reflection minima visualizes the resonance shift of 2.5 nm (Figure \ref{fig2}b,e), which we use to calculate a figure of merit (FOM) that was introduced in our previous work \cite{DamgaardCarstensen2023}: FOM = $\Delta\lambda$/FWHM, where FWHM is full width half maximum. We achieve a FOM of 0.47 and 0.30 for the symmetric and asymmetric gratings, respectively. Both values are an order of magnitude larger than our previous work (0.046), thus verifying the superior performance of this new design. 
Due to the better simulated performance of the symmetric grating (absolute modulation and FOM is higher) and the limitations of our fabrication facilities (not possible to fabricate structures with aspect ratios (height/width) larger than one), we continue with fabrication of the symmetric grating. 

After design optimization, a grating consisting of $N$ = 50 periods was fabricated using electron-beam lithography and the spectrum was measured (Supplementary Figure S3). The measured spectrum shows a very broad and shallow resonance that does not match the simulated spectrum. We conduct a numerical investigation of the connection between the grating size, i.e., the number of unit cell periods, and the resonance shape through numerical simulations of a finite grating with sizes ranging from $N$ = 40 to $N$ = 140 unit cell periods. It is clearly seen that a larger grating gives a deeper and narrower resonance (Supplementary Figure S3). Therefore, we conduct thorough simulations for a finite model with $N$ = 120 unit cell periods. Switching from the infinite to the finite model influences the reflection minimum that is now less deep, meaning the resonant excitation is shifted away from critical coupling (Supplementary Figure S4).

\subsection*{Fabrication and static characterization of metasurface}

\begin{figure}[tb]
	\centering
	\includegraphics{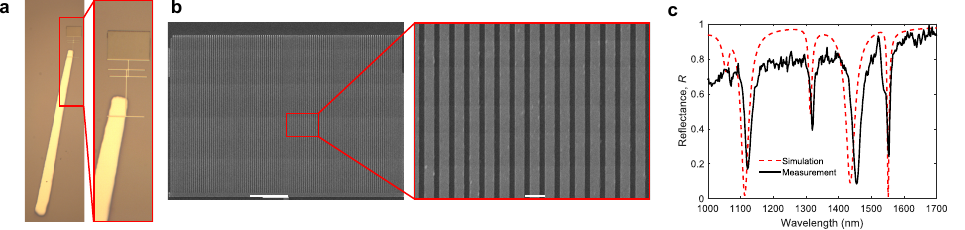}
	\caption{Fabrication and static characterization of qBIC metasurface. \textbf{a} Optical images of a macroscopic electrode and metasurface. \textbf{b} SEM images of the entire grating ($N$ = 120) and a zoom-in view of the fabricated gold ridges. Scale bars represent 10 µm (left) and 1 µm (right). \textbf{c} Measured (black line) and simulated (red dashed line) reflection spectra from the metasurface. Error bars are on the order of the line width. }
	\label{fig3}
\end{figure}

The fabrication process starts by evaporation of large macroscopic electrodes through a shadow-mask (Figure \ref{fig3}a). Afterwards, the gratings are formed based on the design described in Figure \ref{fig2}a,c using the standard nanofabrication technology of electron-beam lithography and lift-off. During this process, the grating is aligned and interfaced with the macroscopic electrode for later application of bias voltages. Scanning electron microscopy images after lift-off show straight evenly-spaced gold ridges with very few defects (Figure \ref{fig3}b). For optical characterization of the static performance, a collimated beam from a supercontinuum source is linearly polarized and inbound on the grating. The reflected light is separated from the incident by a beam splitter, collected by a 20X objective, spatially filtered by a pinhole and collected by a camera or spectrometer (Supplementary Figure S5). The measured and simulated reflection spectra show very similar wavelength dependence in the large wavelength range from 1000-1700 nm for four pronounced reflection minima (Figure \ref{fig3}c). The two deepest and widest reflection minima ($\sim\,$1120 nm and $\sim\,$1460 nm) are Fabry-Perot modes, whereas the two narrow but less deep reflection minima ($\sim\,$1320 nm and $\sim\,$1550 nm) are excitation of guided modes. In this work, we designed around the right-most minimum at 1554 nm, which corresponds to the qBIC resonance discussed in the previous section. 

Ultrafast and ultrathin spatial light modulators are a very desirable ideal application prospect for efficient free-space light modulators. The lateral miniaturization of such devices is directly related to the achievable resolution and device footprint. Under the assumption of the modulator functioning at critical coupling, we calculate the minimal achievable pixel size from the estimated decay length of the optical mode \cite{Weiss2022,DamgaardCarstensen2022}. The loaded Q-factor of our device is $\sim\,$180, which corresponds to an unloaded Q-factor of $Q_U = 2Q_L\approx 360$ at critical coupling, where absorption and scattering losses are equal. The loss per unit length is estimated from the unloaded Q-factor:

\begin{equation}
	\alpha = \frac{4\pi}{Q_U\lambda_{eff}} \approx 0.045 \, \mathrm{\mu m}^{-1}
\end{equation}

where $\lambda_{eff}$ is the effective mode wavelength, and we assume an effective mode refractive index of 2. From the loss per unit length, the propagation length is derived as the inverse: $L_p \approx$ 1/0.045 µm\textsuperscript{-1} = 22 µm. We note that the pixel size of an SLM configuration can be considerably reduced without risking crosstalk due to guided modes. 

\subsection*{Dynamic characterization of metasurface}
\begin{figure}[tb]
	\centering
	\includegraphics{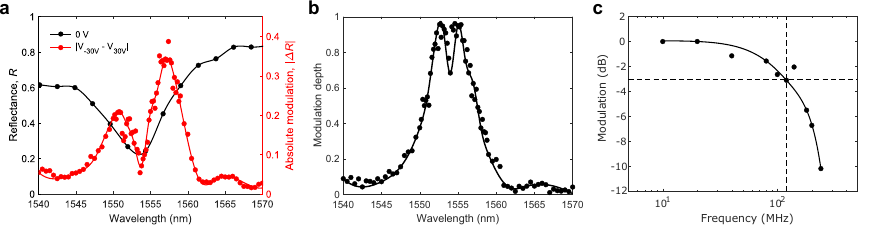}
	\caption{Dynamic characterization of the fabricated tunable qBIC metasurface. \textbf{a} Measured reflection intensity (left axis, black dots and line) and absolute modulation (right axis, red dots and line) vs. wavelength for bias voltages of 0 V and ±30 V, respectively. Measurements are performed using a 1 kHz sinusoidal signal. A moving average of three points has been applied to the absolute modulation data points. \textbf{b} Measured modulation depth vs. wavelength of incident light for bias voltage of $\pm$30 V. \textbf{c} Measured modulation vs. electrical frequency of the applied signal at 1556 nm. The dashed line represents -3 dB. \textbf{a-c} Dots are measurement points and lines are spline approximations to guide the eye. Error bars are on the order of data point sizes. }
	\label{fig4}
\end{figure}

To measure efficient modulator operation, we supply bias voltages to the sample using a function generator and detect reflected signals using a photodetector. By applying a bias voltage of $\pm$30 V, we measure an absolute modulation of up to 35 \% in the wavelength range of 1540 to 1570 nm (Figure \ref{fig4}a). The absolute modulation is larger on the longer wavelength side of the reflection minimum, because the slope of the spectrum is larger. We note that the large spacing between the measurement points of the spectrum are due to the limited resolution of the spectrometer. In addition, a near-perfect modulation depth reaching 95 \% is measured (Figure \ref{fig4}b), which corresponds very well with the simulations (Supplementary Figure S2b). The large contrast between the on- and off-state is visualized in the Supplementary Video S1. As discussed in the previous section, the location of reflection minima can be controlled by changing the angle of incidence of the laser. This effect is investigated in simulations and verified in measurements (Supplementary Figure S1), where we measure the spectrum and the modulation depth for an angle of incidence of 1\textsuperscript{o}. As expected, the reflection minima are less pronounced and therefore the measured modulation depth is also reduced to the range of 30-40 \%. This proves that our metasurface modulator is tunable within a range of more than 30 nm, while maintaining a decent modulation depth. However, the presence of a bandgap in the dispersion relation results in a range of unachievable wavelengths just below the wavelength of the reflection minimum for normal incidence. 

The main advantage of making dynamic metasurfaces utilizing the electro-optic Pockels effect is the inherently fast modulation, which it supports. The electrical bandwidth of our device is measured in the range of 10-250 MHz, and we determine a -3 dB cutoff of 125 MHz (Figure \ref{fig4}c). The limiting factor of the achievable modulation speed is the electrical circuit, and herein primarily the size of the macroscopic electrode, which forms a capacitor with the continuous bottom gold electrode. We verify this by testing another electrode configuration, which is fabricated simultaneously with the grating by the use of electron beam lithography (Supplementary Figure S6). This way, the electrode can be more compact and consequently have a smaller surface area. A -3 dB cutoff frequency of 350 MHz is measured. Further limiting the size of the top electrode or limiting the bottom electrode to only span an area below the patterned grating would allow our metasurface modulator to reach electrical bandwidths of up to 39 GHz estimated for the metasurface pixel size of 22 µm, which is easily supported by the Pockels effect. 

One possible application for high Q factor tunable metasurfaces, is to implement it for switchable phase contrast imaging. The large Q factor in combination with the highly angle-dependent spectral position of the minimum, leads to our metasurface attenuating paraxial rays at wavelengths close to resonance, while not influencing rays incoming at larger angles \cite{Ji2022}. We demonstrate a proof-of-concept of switchable phase contrast imaging of a transparent Poly(methyl methacrylate) (PMMA) stripe, where the edge of the phase object is clearly emphasized at resonance and the contrast is decreased when the metasurface is shifted off resonance by tuning of the applied bias voltage (Supplementary Note 6). 

\section*{Discussion}
To summarize, we have presented an investigation of the occurrence of BIC and qBIC modes in a material configuration consisting of a LN thin-film sandwiched between an optically thick gold back-reflector and a grating of gold nanoridges also functioning as control electrodes. Furthermore, we utilize the findings of this investigation to design, fabricate and experimentally characterize a metasurface for tunable optical free-space intensity modulation. The fabricated electro-optic metasurfaces operate at telecom wavelengths with the modulation depth reaching 95 \% combined with an absolute modulation of 35 \% for a bias voltage of $\pm$30 V within the electrical bandwidth of 125 MHz. In recent years, several free-space intensity modulators utilizing various active materials have been demonstrated (Supplementary Table S1). We believe that the electro-optic LN metasurface configuration presented here is attractive due to its highly efficient performance combined with inherently ultrafast responses, making it superior to those based on phase-change materials or MEMS components, while its exceptional environmental stability contrasts starkly to those based on electro-optic polymers with low glass temperatures. 

Furthermore, highly angle-dependent qBIC resonances realized with this platform open exciting possibilities for ultrafast, electrically tunable phase contrast imaging that allows one to realize electrically tunable phase contrast imaging and, for example, dynamically adjust edge enhancement effects in imaging applications. The estimated pixel size of 22 um is small enough for implementing sophisticated spatial and temporal modulation of radiation at telecom wavelengths similar to what has been demonstrated with the state-of-the-art ITO-based electrically tunable metasurfaces, but with much larger bandwidths \cite{Sisler2024}.  Finally, given the potential bandwidth of 39 GHz estimated for and in combination with the metasurface pixel size of 22 µm, the demonstrated electro-optic metasurfaces promise successful realization of unique optical functions, such as harmonic beam steering and spatiotemporal shaping as well as nonreciprocal operation. Overall, we believe that the demonstrated electro-optic LN metasurfaces based on qBIC resonances open new avenues towards the development of ultrafast, highly efficient and ultrathin flat-optics components for advanced applications involving spatiotemporal control of optical fields.

\section*{Methods}

\subsection*{Simulation}
Simulations are performed in the commercially available finite element software COMSOL Multiphysics, ver. 6.2. Given that the grating design is constant (semi-infinite) in the $y$-direction, all simulations are performed for 2D models. Due to the significant thickness of the chromium adhesion layer (10 nm), and our inability to alter it, we include it in the optical simulations, even though adhesion layers are typically not modelled. Interpolated values are used for the permittivity of chromium \cite{Johnson1974}, LN \cite{Zelmon1997}, and gold \cite{Rakic1998}. For the infinite model simulations we simply calculate one period of the grating with one nanoridge (or two nanoridges for the asymmetric grating) and add periodic boundary conditions to the sidewalls of the model domain. For the finite model, we model the whole grating and truncate the domain with perfectly matched layers to eliminate reflection from boundaries. Wave excitation and measuring is done using ports. An electro-optic simulation is performed in two steps: First, the electric field distribution from an applied DC voltage is determined in an electrostatic simulation. Second, the change of refractive index is calculated from the electric field distribution and the Pockels coefficients of Jazbinšek et al. \cite{Jazbinek2002} (considering only the largest diagonal terms, i.e., $\Delta n_i = -\frac{1}{2}n_i^3r_{iiz}E_z$, with $r_{xxz} = r_{yyz} =$ 10.12 pm/V and $r_{zzz} =$ 31.45 pm/V), after which the optical simulation is conducted with the updated refractive index. For calculation of the dispersion relation and Q factor, we employ the eigenfrequency solver of COMSOL Multiphysics.\\

\subsection*{Fabrication}
The device is fabricated using a combination of nanostenciling and electron beam lithography. 5 nm of titanium and 100 nm of gold is deposited by thermal evaporation through a shadow mask to form macroscopic electrodes. A $\sim\,$200 nm layer of PMMA 950K A4 resist is spin-coated, and the modulator is manually aligned to the macroscopic electrode and exposed using electron beam lithography at 30 kV. The resist is developed, and the modulator is formed by evaporation of 3 nm of titanium and 75 nm of gold followed by liftoff in acetone. The grating array consists of 120 periods, and it measures 90 µm × 60 µm. \\

\subsection*{Electro-optic characterization}
The sample was mounted on a homemade sample holder, with an attached commercial probe (GGB, Model 40A-GSG-750, working range: DC - 40 GHz) used for connection to the upper electrode (the grating), while to connect to the bottom electrode a conductive paste is applied to the edge of the sample. The sample holder was mounted on a 3D stage. For characterization of the passive device, a collimated supercontinuum laser beam (NKT Photonics SuperK Extreme) is used in combination with a spectrometer to get the full spectral response, whereas for the characterization of the active device, a collimated low-power continuous-wave laser beam from a tunable telecom laser (New Focus Venturi) is used. The polarization of the incident light is controlled by a Glan-Thompson polarizer, and a beam splitter is used for the sample illumination, placed between the sample and the objective in the collection part, to maximally resemble the plane wave incidence. The reflected light is collected by a 20X objective, spatially filtered with a pinhole to reduce the background noise and further carried to the photodetector or spectrometer. The measurements of the modulation depth are performed at relatively low frequencies ($\sim\;$200 kHz), with the modulating signals supplied by a homemade function generator and measured with an oscilloscope (bandwidth of 200 MHz). Electrical bandwidth is determined using a high-speed photodetector and RF spectrum analyzer with high sensitivity. \\

\noindent \textbf{Research funding:} C.D.-C. and S.I.B. acknowledge financial support from Villum Fonden (Award in Technical and Natural Sciences 2019). T.Y. acknowledges the support from the Center for Polariton-driven Light-Matter Interactions (POLIMA) funded by the Danish National Research Foundation (Project No. DNRF165). \\

\noindent \textbf{Conflict of interest statement:} The authors declare no conflicts of interest. \\

\bibliographystyle{ieeetr} 
\bibliography{AMJ}

\end{document}